\documentstyle[12pt,axodraw,epsfig,psfig]{article}
\begin{document}
\baselineskip 0.6cm
\newcommand{\gsim}{ \mathop{}_{\textstyle \sim}^{\textstyle >} }
\newcommand{\lsim}{ \mathop{}_{\textstyle \sim}^{\textstyle <} }
\newcommand{\vev}[1]{ \left\langle {#1} \right\rangle }
\newcommand{\bra}[1]{ \langle {#1} | }
\newcommand{\ket}[1]{ | {#1} \rangle }
\newcommand{\EV}{ {\rm eV} }
\newcommand{\KEV}{ {\rm keV} }
\newcommand{\MEV}{ {\rm MeV} }
\newcommand{\GEV}{ {\rm GeV} }
\newcommand{\TEV}{ {\rm TeV} }
\def\diag{\mathop{\rm diag}\nolimits}
\def\Spin{\mathop{\rm Spin}}
\def\SO{\mathop{\rm SO}}
\def\O{\mathop{\rm O}}
\def\SU{\mathop{\rm SU}}
\def\U{\mathop{\rm U}}
\def\Sp{\mathop{\rm Sp}}
\def\SL{\mathop{\rm SL}}
\def\tr{\mathop{\rm tr}}


\begin{titlepage}

\begin{flushright}
UT-978
\end{flushright}

\vskip 2cm
\begin{center}
{\large \bf  Proton Decay in the Semi-Simple Unification}

\vskip 1.2cm
Masaaki Fujii and  T.~Watari

\vskip 0.4cm
{\it Department of Physics, University of Tokyo, \\
         Tokyo 113-0033, Japan}\\

\vskip 1.5cm
\abstract{Semi-simple unification is one of a model which naturally 
solves two difficulties in the supersymmetric grand unification theory: 
doublet-triplet splitting problem and suppression of dimension 5 
proton decay. We analyzed the dimension 6 proton decay of this model using
perturbative analysis at the next-to-leading order. The life time of proton 
is $3 \times 10^{34}$ - $10^{35}$ years for wide range of SUSY breaking 
parameters, and there is an intriguing possibility of observing proton
decay signals in the next-generation water \v{C}erenkov detectors 
such as Hyper-Kamiokande and TITAND. 
Several uncertainties in this prediction are also discussed.}

\end{center}
\end{titlepage}


{\bf Introduction}

Supersymmetric(SUSY) SU(5)$_{\rm GUT}$ grand unification theories(GUT's) 
are supported by the approximate SU(5)$_{\rm GUT}$ unification 
at around $10^{16} \GEV$ of the three gauge coupling constants 
of the minimal supersymmetric standard model(MSSM)\cite{LL}.
However, the conceptual beauties of the GUT's\cite{GUT} and 
such a phenomenological success are not more than 
indirect evidences, and it would be the proton decay signals 
that makes us believe the GUT's in nature.

The minimal SU(5)$_{\rm GUT}$ model predicts proton decay 
through dimension 5 operators\cite{dim5}, and is now 
almost excluded\cite{Murayama} by experimental 
bounds such as  $\tau (p \rightarrow K^+ \bar{\nu}) \gsim 
6.7 \times 10^{32} {\rm yr}.$(90 \% C.L.)\cite{SKdim5}.
Therefore, we have to analyze the proton decay in an extended model
in which 
those operators are suppressed or absent.
Predictions on the proton decay through dimension 6 operators 
severely depend on how a model is extended\footnote{
For example, in some type of model~\cite{Witten}, 
the dimension 6 operators are not induced by the GUT gauge boson exchange.};
the life time of the proton depends on the fourth power of the mass 
of SU(5)$_{\rm GUT}$-off-diagonal gauge boson (GUT gauge boson), 
and hence on the detailed spectrum of the model around the GUT scale.
Therefore, an analysis on the proton decay has to be based on a
phenomenologically reliable model of the GUT's.
 
The doublet-triplet splitting problem\cite{d-t} and 
suppression of the dimension 5 proton decay operators had been 
the two major obstacles in model buildings of the GUT's. 
The semi-simple unification\cite{ss1,ss2,ss3} is a model
that solves these two problems in a natural way.
In this letter, we calculate the proton decay rate in this model.
The proton decay is relatively fast in this model, 
whose reason will be clear in the text. 
We restrict ourselves in parameter region of the model 
where a perturbative analysis is valid. 
As a result of a full next-to-leading order analysis\cite{HMY}, 
we found that the mass of the 
GUT gauge boson can be determined, 
and that the resulting life time of proton does not depend on 
SUSY breaking parameters so much: 
the life time is  $\tau \simeq (3-10) \times 10^{34} {\rm{yr}}$.
Various sources of uncertainties in this prediction 
are summarized at the end of this letter.
This result means that the proton decay is generically detectable 
in the next-generation water \v{C}erenkov detectors such as 
Hyper-Kamiokande and TITAND~\cite{HyperK,TITAND}.

{\bf Brief Review of the Semi-Simple Unification Model}

We briefly review the semi-simple unification model that uses 
$\SU(5)_{\rm GUT}  \times \U(3)_{\rm H}$ gauge group.
This gauge group is directly broken down to the
$\SU(3)_{\rm C} \times \SU(2)_{\rm L} \times \U(1)_{\rm Y}$ of the MSSM.
Quark and lepton (${\bf 5^* + 10}$) and Higgs ($H^i({\bf 5}) +
\bar{H}_i({\bf 5}^*)$) supermultiplets are singlets of 
the $\U(3)_{\rm H}$ gauge group and transform under the $\SU(5)_{\rm GUT}$ 
as in the standard $\SU(5) _{\rm GUT}$ model. 
Some more fields are required for the GUT symmetry breaking:
$X^{\alpha}_{\;\;\beta}(\alpha ,\beta =1,2,3)$ transforming 
as (${\bf 1}$,{\bf adj.}=${\bf 8}+{\bf 1}$) under 
the $\SU(5)_{\rm GUT}\times \U(3)_{\rm H}$ gauge 
group, and $Q^{\alpha}_{\;i}(i=1,\cdots,5)+Q^{\alpha}_{\;6}$
and $\bar{Q}^i_{\;\alpha}(i=1,\cdots,5)+\bar{Q}^6_{\;\alpha}$
transforming as (${\bf 5}^*$ + ${\bf 1}$,${\bf 3}$) and 
(${\bf 5 + 1}$,${\bf 3}^*$).
Indices $\alpha$ and $\beta$ are for the $\U(3)_{\rm H}$ and $i$ 
for the $\SU(5)_{\rm GUT}$. $X^\alpha_{\;\;\beta}$ is also expressed as 
$X^c (t_c)^\alpha_{\;\;\beta}(c=0,1-8)$, where $t^a(a=1-8)$ are 
Gell'mann matrices\footnote{A normalization condition 
$\tr(t_a t_b) = \delta_{ab}/2$ is understood. Note that the
normalization of the following $t_0$ is determined so that 
it also satisfies $\tr(t_0t_0)=1/2$.} and 
$t_0 \equiv {\bf 1}_{3 \times 3}/\sqrt{6}$.
Superpotential is given as\cite{ss3} 
\begin{eqnarray}
W &=& 
\sqrt{2} \lambda_{\rm 3H} \bar{Q}^i_{\;\alpha} X^a(t_a)^{\alpha}_{\;\beta}
                             Q^{\beta}_{\;i} 
+ \sqrt{2} \lambda_{\rm 3H}' \bar{Q}^6_{\;\alpha} X^a(t_a)^{\alpha}_{\;\beta}
                             Q^{\beta}_{\;6}   \nonumber \\
& & 
+ \sqrt{2} \lambda_{\rm 1H} \bar{Q}^i_{\;\alpha} X^0(t_0)^{\alpha}_{\;\beta}
                             Q^{\beta}_{\;i} 
+ \sqrt{2} \lambda_{\rm 1H}' \bar{Q}^6_{\;\alpha} X^0(t_0)^{\alpha}_{\;\beta}
                             Q^{\beta}_{\;6}  \nonumber \\
& &
- \sqrt{2}\lambda_{\rm 1H}  v^2 X^\alpha_{\;\alpha}    \label{eq:super}  \\
& & + h' \bar{H}_i \bar{Q}^i_{\;\alpha} Q^{\alpha}_{\;6} 
      + h \bar{Q}^6_{\;\alpha} Q^{\alpha}_{\;i}H^i     \nonumber  \\
& & + y_{\bf 10} {\bf 10} \cdot {\bf 10} \cdot H 
      + y_{\bf 5^*} {\bf 5}^* \cdot {\bf 10} \cdot \bar{H} + \cdots,
\nonumber
\end{eqnarray}
where 
the parameter $v$ is of order of the GUT scale, 
$y_{\bf 10},y_{\bf 5^*}$ are Yukawa coupling constants 
of the quarks and leptons, and $\lambda_{\rm 3H},\lambda_{\rm 3H}',
\lambda_{\rm 1H},\lambda_{\rm 1H}',h',h$ are dimensionless coupling constants.
One can see that the above superpotential has ${\bf Z}_4$ R symmetry
under a charge assignment given in Table \ref{tab:Z4-charge}, and this
symmetry forbids enormous mass term $W = \bar{H} H$ for the 
Higgs doublets\footnote{$\mu$-term can be obtained through the 
Giudice-Masiero mechanism\cite{GM}.}.
The bifundamental representation 
$Q^{\alpha}_{\;i}$ and $\bar{Q}^i_{\;\alpha}$ acquire vacuum expectation
value, $\vev{Q^\alpha_{\;\;i}}= v \delta^\alpha_i$ and 
$\langle$ $\bar{Q}^i_{\;\;\alpha}$ $\rangle$ $= v \delta^i_\alpha$,  
because of the second and the third line in (\ref{eq:super}).
Then, the mass terms of the colored Higgs multiplets arise from 
the fourth line in (\ref{eq:super}) in the GUT-breaking vacuum. 
The ${\bf Z}_4$ R symmetry is not broken even after the GUT symmetry 
is broken. One can also see that this ${\bf Z}_4$ R symmetry forbids 
the dangerous dimension 5 proton decay operators 
$W = {\bf 10}\cdot{\bf 10}\cdot{\bf 10}\cdot{\bf 5}^*$. 

{\bf Naive Estimation}

\begin{figure}[h!]
 \centerline{ {\psfig{figure=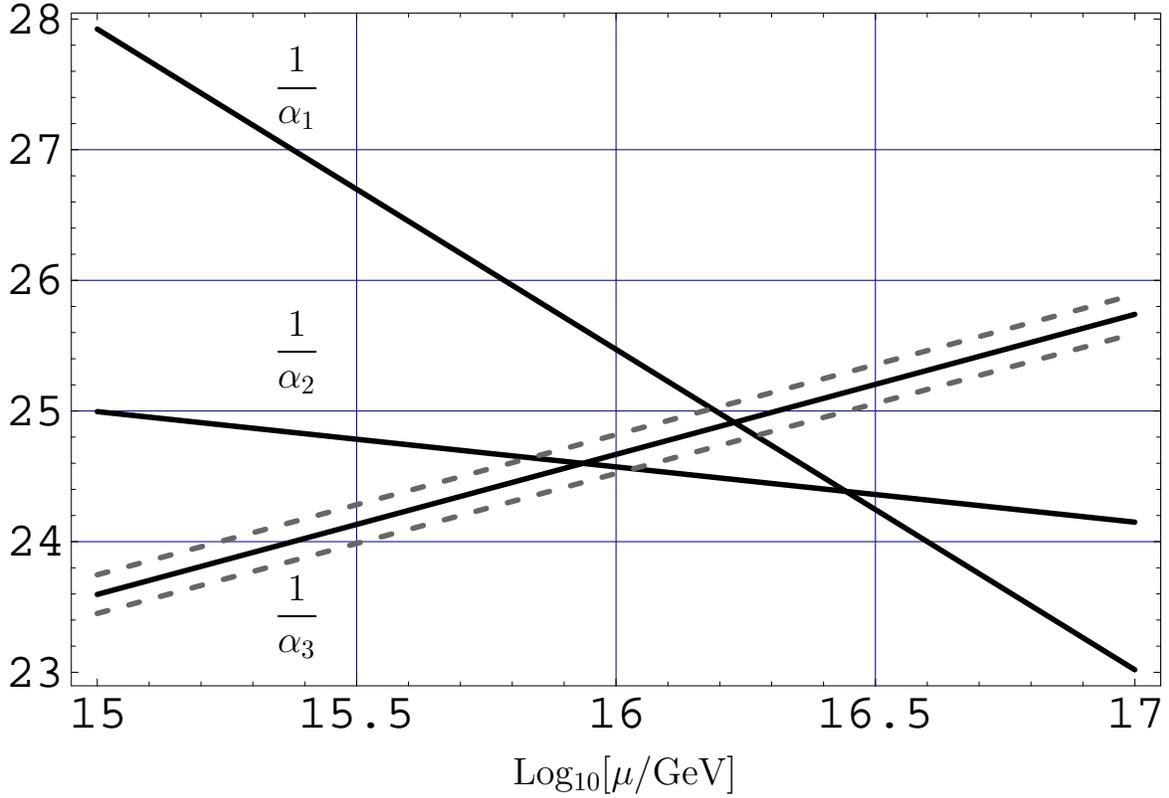,height=10cm}} }
  \begin{picture}(0,0)
\large
  \put(120,260){$\displaystyle{\frac{1}{\alpha_{1}}}$}
  \put(120,160){$\displaystyle{\frac{1}{\alpha_{2}}}$}   
  \put(120,60){$\displaystyle{\frac{1}{\alpha_{3}}}$}

\put(210,0){Log$_{10}$[$\mu$/GeV]}
\normalsize
\end{picture}
 \caption{Approximate SU(5)$_{\rm GUT}$ relation between 
the three MSSM gauge coupling constants and deviation from it.
1$\sigma$ error bar of the QCD coupling are also described. 
SUSY threshold corrections are calculated using the spectrum of 
mSUGRA model with $m_0=250 \GEV$, $M_{1/2}=500 \GEV$, $A_0 = 0$ and
 $\tan \beta = 10$. The sign of $\mu$-term is taken to be negative. }
 \label{fig:triangle}
\end{figure}

At tree level, the gauge coupling constants of the 
$\SU(3)_{\rm C} \times \SU(2)_{\rm L} \times \U(1)_{\rm Y}$ are given 
in terms of those of the $\SU(5)_{\rm GUT} \times \U(3)_{\rm H}$ as :
\begin{equation}
 \left(\frac{1}{\alpha_3} \equiv \frac{1}{\alpha_C} \right) 
    = \frac{1}{\alpha_{\rm GUT}} + \frac{1}{\alpha_{\rm 3H}},
\label{eq:tree3}
\end{equation}
\begin{equation}
 \left(\frac{1}{\alpha_2} \equiv \frac{1}{\alpha_L} \right)
    =  \frac{1}{\alpha_{\rm GUT}} \quad \quad \quad
\end{equation}
and 
\begin{equation}
\left(\frac{1}{\alpha_1} \equiv \frac{3/5}{\alpha_Y} \right)
    =  \frac{1}{\alpha_{\rm GUT}} + \frac{2/5}{\alpha_{\rm 1H}},
\label{eq:tree1}
\end{equation} 
where $\alpha_C, \alpha_L, \alpha_Y, \alpha_{\rm GUT}, 
\alpha_{\rm 3H}$ and $\alpha_{\rm 1H}$ are fine structure constants 
of the three MSSM gauge groups, $\SU(5)_{\rm GUT}$, 
$\SU(3)_{\rm H}$ and $\U(1)_{\rm H}$\footnote{The normalization of the
$\U(1)_{\rm H}$-generator is determined so that $Q,\bar{Q}$ have charge 
$\pm 1/\sqrt{6}$.}, respectively. 
The approximate SU(5)$_{\rm GUT}$ relation and deviation form it
(Fig.\ref{fig:triangle}) are naturally explained through 
the above equations if $1/\alpha_{\rm GUT} \sim 24$ and 
$1/\alpha_{3H} \lsim 1, 1/\alpha_{1H} \lsim 2.5$. 
At the same time, we notice that the ``GUT scale'', an energy scale at
which Eqs.(\ref{eq:tree3}-\ref{eq:tree1}) hold, lies lower than the 
$\alpha_1$ - $\alpha_2$ unification scale $M_{1-2}$
and  higher than the 
$\alpha_2$ - $\alpha_3$ unification scale $M_{2-3}$. 
Therefore, the decay rate of proton is expected to be enhanced 
compared with the rate using the $M_{1-2}$ as the GUT gauge boson mass.

At one-loop order, the gauge coupling of the $\U(1)_{\rm H}$ 
runs\footnote{The one-loop $\beta$-function of the $\SU(3)_{\rm H}$ 
coupling is 0.} asymptotic non-free :
\begin{equation}
\frac{\partial }{\partial \ln \mu} 
\left( \frac{1}{\alpha_{\rm 1H}}(\mu) \right) 
= - \frac{6}{2 \pi}.
\label{eq:1H-1loop}
\end{equation}
There are two important remarks here.
First, the cut-off scale $\Lambda$ of this model exists below the Planck 
scale; $1/\alpha_{\rm 1H}$ should be positive even at the $\Lambda$.
The constraint $1/\alpha_{\rm 1H} \lsim 2.5$ at the ``GUT scale'' allows 
the $\Lambda$ to be higher than the ``GUT scale'' by one order 
of magnitude or a little bit more, and this is much enough to 
justify the field theoretical description of the GUT symmetry breaking
and to accommodate all the GUT spectrum below the cut-off scale
$\Lambda$. This $\Lambda$ lies around $10^{17} \GEV $ or a little 
higher, though it may be below $10^{18} \GEV$.
Secondly, the IR-free (asymptotic non-free) behavior of 
the $\U(1)_{\rm H}$ coupling leads to
\begin{equation}
\frac{1}{\alpha_{\rm 3H}} \ll \frac{1}{\alpha_{\rm 1H}}
\label{eq:lefty}
\end{equation}
at the ``GUT scale'' under an assumption
\begin{equation}
 \frac{1}{\alpha_{\rm 3H}}(\Lambda) \simeq \frac{1}{\alpha_{1H}}(\Lambda).
\label{eq:U(3)-relation}
\end{equation}
This $\U(3)_{\rm H}$-relation at $\Lambda$ is quite natural if 
there is $\U(3)_{\rm H}$-structure in a fundamental theory~\cite{IWY}.
Then, as a consequence of the relation Eq.(\ref{eq:lefty}), we notice 
that the ``GUT scale'' is closer to the $M_{2-3}$ rather than to the
$M_{1-2}$ since $(1/\alpha_1 - 1/\alpha_2) > (1/\alpha_3 - 1/\alpha_2)$
there, and hence the proton decay is relatively fast.

{\bf Threshold Corrections at the GUT Scale}

In the analysis at the next-to-leading order, one-loop threshold
corrections of the GUT model are also taken into account.
The three MSSM gauge coupling constants just below the GUT scale
are expressed in terms of the gauge coupling constants and 
various masses in the spectrum of the GUT model, including
the mass of the GUT 
gauge boson, $M_v$. 
Particle spectrum around the GUT scale is summarized in 
Table \ref{tab:spec}. 
Explicit expressions of the MSSM gauge coupling constants 
are given as follows :
\begin{eqnarray}
\hspace{-0.4cm}&&\frac{1}{\alpha_3}(\mu) = 
\frac{1}{\alpha_{\rm GUT}}(\Lambda) + \frac{1}{\alpha_{\rm 3H}}(\Lambda) +
\frac{-3}{2\pi}\ln\left(\frac{\Lambda}{\mu}\right)+
\frac{ 1}{2\pi}\ln\left(\frac{\Lambda^2}{M_{\bar{c}}M_c}\right)- 
\frac{ 4}{2\pi}\ln\left(\frac{\Lambda}{M_{v}}\right)+
\frac{ 6}{2\pi}\ln\left(\frac{M_{8v}}{M_{8c}}\right) ,
\label{eq:full-matching3}\nonumber \\
\\
&&\frac{1}{\alpha_2}(\mu) = 
\frac{1}{\alpha_{\rm GUT}}(\Lambda) \quad\qquad\quad\;\;  +
\frac{ 1}{2\pi}\ln\left(\frac{\Lambda}{\mu}\right)
\qquad \qquad \qquad \qquad\;\; - 
\frac{ 6}{2\pi}\ln\left(\frac{\Lambda}{M_{v}}\right) ,
\label{eq:full-matching2}\\
\nonumber\\
&&\frac{1}{\alpha_1}(\mu) = 
\frac{1}{\alpha_{\rm GUT}}(\Lambda) + 
\frac{2/5}{\alpha_{\rm 1H}}(\Lambda) +
\frac{33/5}{2\pi}\ln\left(\frac{\Lambda}{\mu}\right)+
\frac{2/5}{2\pi}\ln\left(\frac{\Lambda^2}{M_{\bar{c}}M_c}\right)- 
\frac{10}{2\pi}\ln\left(\frac{\Lambda}{M_{v}}\right),
\qquad \qquad \qquad \label{eq:full-matching1}
\end{eqnarray}
where $\mu$ is a renormalization point, which is taken to be 
just below the GUT scale,
$M_v$, $M_c$, $M_{\bar{c}}$, $M_{8v}$, $M_{8c}$ are masses of particles 
around the GUT scale (see Table \ref{tab:spec}) and 
$\alpha_{\rm GUT,3H,1H}(\Lambda)$ are fine structure constants of
the gauge groups $\SU(5)_{\rm GUT},\SU(3)_{\rm H}$ and $\U(1)_{\rm H}$,
respectively, at the cut-off scale $\Lambda$.  

In general, it is impossible to determine the $M_v$ if  
GUT models have more than three parameters.
However, it is not necessarily the case in the semi-simple 
unification model.
Threshold corrections in 
Eqs.(\ref{eq:full-matching3}-\ref{eq:full-matching1}) is simplified
considerably under two reasonable assumptions. One is 
the $\U(3)_{\rm H}$-relation Eq.(\ref{eq:U(3)-relation}) and 
the other is ${\cal N}=2$-SUSY-relation:
\begin{equation}
 g_{\rm 1H} \simeq  \lambda_{\rm 1H}( \sim \lambda_{\rm 1H}'), \qquad \qquad
 g_{\rm 3H} \simeq  \lambda_{\rm 3H}( \sim \lambda_{\rm 3H}'). 
\label{eq:N=2-relation}
\end{equation}
Under the latter condition, a large threshold correction from 
the massive $\SU(3)_{\rm H}$ vector multiplet\footnote{Note that 
$M_{8v} \sim 10 \cdot M_v$, and hence the threshold correction 
is large.} is almost canceled by its ${\cal N}=2$-partner, 
the $\SU(3)_{\rm C}$-${\bf adj.}$ chiral multiplets, 
since $M_{8v} \simeq M_{8c}$.
Now that the threshold corrections form the $\SU(3)_{\rm C}$-${\bf adj.}$
multiplets decouple from
Eqs.(\ref{eq:full-matching3}-\ref{eq:full-matching1}),
we are left only with two threshold corrections: one from 
the  massive vector multiplet of the GUT gauge boson
and the other from colored Higgs chiral multiplets. 
Then, one can easily see that three combinations,
\begin{equation}
  \frac{1}{\alpha_{\rm GUT}}(\Lambda), \quad 
  \ln\left(\frac{\Lambda}{M_v}\right) {\rm ~~and~~} 
  \frac{1}{\alpha_{\rm 3H}}(\Lambda) +
  \frac{ 1}{2\pi}\ln\left(\frac{\Lambda^2}{M_{\bar{c}}M_c}\right), 
\end{equation}
are determined in terms of the values to be put in the LHS's and 
deviation from the $\U(3)_{\rm H}$-relation and the ${\cal
N}=2$-relation, once the cut-off scale $\Lambda$ is fixed.

In particular, the GUT gauge boson mass is given by
\begin{equation}
M_v = \sqrt{\frac{\mu^3}{\Lambda}} \exp \left( 
- \frac{2\pi}{24}\left(
    \frac{2}{\alpha_3} + \frac{3}{\alpha_2} - \frac{5}{\alpha_1}
                  \right)(\mu)               \right)
\sqrt{\frac{M_{8v}}{M_{8c}}} 
\exp \left( - \frac{2 \pi}{12}
    \left(\frac{1}{\alpha_{\rm 1H}}-\frac{1}{\alpha_{\rm 3H}}\right)(\Lambda)
     \right).
\label{GUT-mass}
\end{equation}
The last two factors show how the result is changed due to the deviation 
from the assumptions we made. 
$\Lambda^{-1/2}$-dependence is a direct consequence of the one-loop
running of the $\alpha_{\rm 1H}$ in Eq.(\ref{eq:1H-1loop}), and this 
negative power dependence implies that 
this gauge boson mass is generically light.
The life time of proton through this GUT 
gauge boson exchange is given in terms of the $M_v$ as\cite{HMY}
\begin{equation}
\tau(p \rightarrow \pi^0 e^+) \simeq 0.61 \times 10^{35} \times
\left(\frac{M_v}{10^{16}\GEV}\right)^4 
\left(\frac{1}{24 \alpha_{\rm GUT}(M_v)}\right)^2
\left(\frac{0.15(\GEV)^2}{|W|}\right)^2 {\rm yr.},
\end{equation}
where $W$ is a hadron matrix element calculated with lattice quenched
QCD\cite{lattice}.

{\bf Threshold Corrections at the Weak Scale and Two-Loop Running}

In order to determine the precise value of the GUT gauge boson mass
by using (\ref{GUT-mass}), we must accurately 
determine the three MSSM gauge coupling constants at the GUT scale.
For this purpose, we take full one-loop threshold
corrections at the weak scale into account 
for the three gauge coupling constants 
and top-  and bottom-Yukawa coupling constants by following
the method in Ref.~\cite{BPMZ}, and use two-loop renormalization group(RG)
equations.
For illustration, let us briefly review the procedures which
we adopt in this letter.
The conventions of SUSY breaking parameters and of the 
sign of the $\mu$-term are the same as those in Ref.~\cite{BPMZ}.

The SUSY threshold corrections to fix the $\overline{{\rm{DR}}}$ coupling
$\alpha_{3}(M_{Z})$ is very simple and the result is as follows:
\begin{equation}
\alpha_{3}(M_{Z})=\frac{\alpha_{3}(M_{Z})_{\overline{\rm{MS}}}}{1-\Delta \alpha_{3}},
\end{equation}
where
\begin{equation}
\Delta \alpha_{3}=\frac{\alpha_{3}(M_{Z})_{\overline{\rm{MS}}}}{2 \pi}
\left[\frac{1}{2}-\frac{2}{3}{\rm{ln}}\left(\frac{m_{t}}{2}\right)
-2{\rm{ln}}\left(\frac{m_{\tilde{g}}}{M_{Z}}\right)
-\frac{1}{6}\sum_{\tilde{q}}\sum_{i=1}^{2}{\rm{ln}}\left(
\frac{m_{\tilde{q}_{i}}}{M_{Z}}
\right)  \right].
\label{eq:SUSY-corr-QCD}
\end{equation} 
Here, $M_{Z}=91.188\GEV$ is the Z-boson pole mass and 
we take $\alpha_{3}(M_{Z})_{\overline{\rm{MS}}}=0.118(2)$ \cite{PD}.
The summation with $\tilde{q}$ runs over all the six squark flavors, and
the constant contribution $1/2$ in Eq.(\ref{eq:SUSY-corr-QCD}) is
necessary when the coupling is translated from the 
$\overline{\rm MS}$ scheme to the $\overline{\rm DR}$ scheme.

Because of the breaking effects of the SU$(2)_{\rm{L}}$ gauge group,
the determinations of the $\overline{\rm{DR}}$ gauge coupling constants 
$\alpha_{\rm Y}(M_{Z})$ and $\alpha_{\rm L}(M_{Z})$ are much more complicated. 
First, we calculate the $\overline{\rm{DR}}$ electromagnetic coupling
constant, $\alpha (M_{Z})$. The explicit formula is given by
\begin{equation}
\alpha (M_{Z})= \frac{\alpha_{\rm{em}}}{1-\Delta \alpha},
\qquad \alpha_{{\rm{em}}}=\frac{1}{137.036},
\end{equation}
where
\begin{eqnarray}
&&\Delta \alpha=0.0682\pm0.0007\nonumber\\
&&\hspace{0.8cm}-\frac{\alpha_{\rm{em}}}{2\pi}
\left[
\begin{array}{lll}\displaystyle{
-7{\rm{ln}}\left(\frac{M_{W}}{M_{Z}}\right)
+\frac{16}{9}{\rm{ln}}\left(\frac{m_{t}}{M_{Z}}\right)
+\frac{1}{3}{\rm{ln}}\left(\frac{m_{H^+}}{M_{Z}}\right)+\frac{4}{9}
\sum_{u}\sum_{i=1}^{2}{\rm{ln}}\left(\frac{m_{\tilde{u}_{i}}}{M_{Z}}\right)}\\
\displaystyle{
+\frac{1}{9}\sum_{d}\sum_{i=1}^{2}{\rm{ln}}\left(
\frac{m_{\tilde{d}_{i}}}{M_{Z}}\right)
+\frac{1}{3}\sum_{e}\sum_{i=1}^{2}{\rm{ln}}\left(
\frac{m_{\tilde{e}_{i}}}{M_{Z}}\right)
 +\frac{4}{3}\sum_{i=1}^{2}{\rm{ln}}\left(
\frac{m_{\tilde{\chi}^{+}}}{M_{Z}}\right)
}
\end{array}
\right].      
\end{eqnarray}
Here, $\sum_{u}$ denotes a sum over $u,\;c,\;t,$ and similarly for 
$\sum_{d}$ and $\sum_{e}$. The numerical values appearing in the 
above expression includes the two-loop QED and QCD corrections 
in Ref.~\cite{FKS},
as well as the five-flavor contributions in Ref.~\cite{EJ}.

Next, we need to fix the 
$\overline{\rm{DR}}$ weak mixing angle $\theta_{\rm{ew}}$ to derive the 
$\overline{\rm{DR}}$ gauge coupling constants, $\alpha_{\rm Y}(M_{Z})$ and
$\alpha_{\rm L}(M_{Z})$. 
The formula to get the $\overline{\rm{DR}}$ weak mixing angle is given by
\begin{eqnarray}
{\rm{cos}}^{2}(\theta_{\rm{ew}}){\rm{sin}}^{2}(\theta_{\rm{ew}})
&=&\frac{\pi \alpha(M_{Z})}{\sqrt{2}M_{Z}^{2} G_{\mu}(1-\Delta r)}\;,\\
\nonumber\\
\Delta r&=&\rho \frac{\Pi_{WW}^{T}(0)}{M_{W}^{2}}-{\cal{R}}e
\frac{\Pi_{ZZ}^{T}(M_{Z}^{2})}{M_{Z}^{2}}+\delta_{VB}\;,
\end{eqnarray}
where $M_{W}=80.419 \GEV$ is the W-boson pole mass, $\rho$ is defined as 
$\rho\equiv M_W^2 /( {\rm{cos}}^{2}(\theta_{\rm{ew}}) M_Z^2 )$, 
$G_{\mu}=1.16639\times 10^{-5}\GEV^{-2}$ is the Fermi constant, 
and $\delta_{VB}$ denotes the nonuniversal vertex and box diagram 
corrections.
The explicit formulae to calculate the quantities given in the 
above expressions and the $\overline{\rm{DR}}$ Yukawa
coupling constants are all given in Ref.~\cite{BPMZ}.

Taking $\alpha_{{\rm{em}}}$, $\alpha_{3}(M_{Z})_{\overline{\rm{MS}}}$, 
the quark and lepton masses, 
and SUSY particle masses as inputs, we calculate all the three gauge
coupling constants and top- and bottom-Yukawa coupling constants
in $\overline{\rm{DR}}$ scheme at the Z-boson pole mass with full one-loop 
threshold corrections.
With these values and tree level tau-Yukawa coupling,
we use the two-loop RG equations 
to obtain the gauge coupling constants at the GUT scale.
For the Yukawa coupling constants, we use one-loop RG equations.

In this letter, 
we adopt the central values given in Ref.~\cite{PD} for the masses 
of vector bosons, quarks and leptons\footnote{Neutrino masses 
are set to be zero.}. 
As for the mass spectrum of the SUSY particles and light Higgs particle, 
we take the 
values calculated by the {\it{SOFTSUSY}} code~\cite{SSUSY} with mSUGRA 
boundary conditions for demonstration.\footnote{
We greatly thank K.Suzuki for this task.}
By using these input values with mSUGRA boundary conditions,
we also confirm that the unification-scale correction $\epsilon_{g}$
\begin{equation}
\alpha_{3}(M_{1-2})=\alpha_{1}(M_{1-2})(1+2 \epsilon_{g}) ,
\end{equation}
at the $\alpha_{1}$--$\alpha_{2}$ unification scale $M_{1-2}$ is 
quite consistent with the result given in Ref.~\cite{BPMZ}. 
  
{\bf Conclusion}

Now, we can estimate the proton life time for various SUSY particle spectra.
We neglect, for the moment, possible two uncertainties expressed by the
last two factors in (\ref{GUT-mass}) coming from the deviation from 
${\cal N}=2$-relation and $\U(3)_{\rm H}$-relation.
Effects of such deviations are discussed later.
Here, we also set the cut-off scale to be $10^{17}\GEV$;
In most part of SUSY breaking parameter space, 
the three gauge coupling constants unify approximately at around 
$10^{16}\GEV$ and hence the cut-off scale $\Lambda$ is expected to 
be no less than $10^{17}\GEV$. 
Therefore, we obtain a conservative upper bound of the proton life time,
using the lowest cut-off scale (see (\ref{GUT-mass})).

\begin{figure}[tbp]
 \centerline{ {\psfig{figure=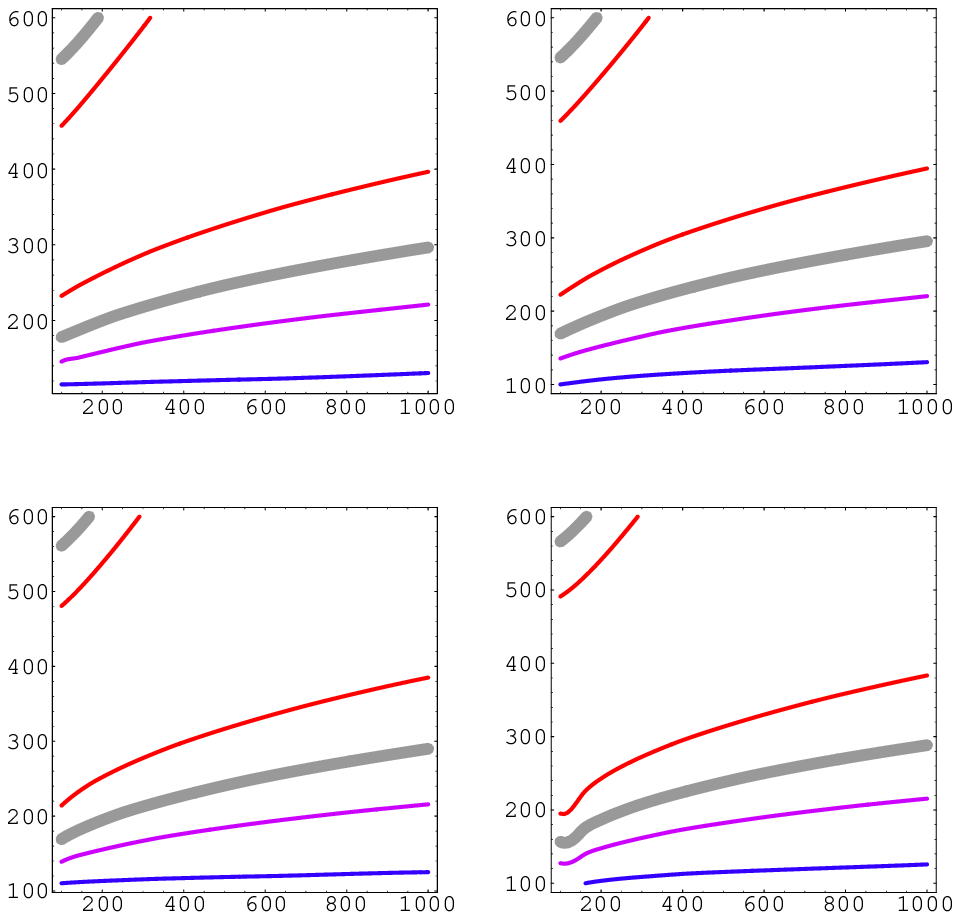,height=16cm}} }
  \begin{picture}(0,0)
\small  
 \put(-20,130){$M_{1/2}[\mbox{GeV}]$}
   \put(100,20){$m_{0}[\mbox{GeV}]$} 
 \put(-20,355){$M_{1/2}[\mbox{GeV}]$}
\put(335,20){$m_{0}[\mbox{GeV}]$} 
\put(90,125){$\bf{5\times10^{34}}$}
\put(65,190){$\bf{5\times10^{34}}$}
\put(185,88){$\bf{10^{35}}$}
\put(80,228){${\rm{tan}}\beta=30,\;A_{0}=0$}
\put(90,355){$\bf{5\times10^{34}}$}
\put(65,410){$\bf{5\times10^{34}}$}
\put(185,318){$\bf{10^{35}}$}
\put(80,455){${\rm{tan}}\beta=10,\;A_{0}=0$}
\put(315,125){$\bf{5\times10^{34}}$}
\put(290,190){$\bf{5\times10^{34}}$}
\put(405,90){$\bf{10^{35}}$}
\put(290,455){${\rm{tan}}\beta=10,\;A_{0}=-300\GEV$}
\put(315,355){$\bf{5\times10^{34}}$}
\put(290,410){$\bf{5\times10^{34}}$}
\put(405,318){$\bf{10^{35}}$}
\put(290,228){${\rm{tan}}\beta=30,\;A_{0}=-300\GEV$}
\normalsize
\end{picture}
 \caption{Contour plots of the proton life time in $m_{0}-M_{1/2}$
 plane for $\mu<0$ cases.  SUSY threshold corrections are calculated
 using mSUGRA sparticle spectrum with universal boundary conditions
 $m_{0}$, $M_{1/2}$, $A_{0}$ at the GUT scale. As for
 $({\rm{tan}}\beta,\; A_{0}[\mbox{GeV}])$, we take them to be $(10,0)$,
 $(10,-300)$, $(30,0)$, $(30,-300)$, respectively as you can see from each
 figure. Solid lines correspond to the contours of the proton 
life time, $5\times 10^{34}$yr., $7\times10^{34}$yr., $10^{35}$yr.,
$2\times 10^{35}$yr. from in to out, respectively. 
Some of them are explicitly denoted in each figure. 
 } \label{fig:myunegative}
\end{figure}

\begin{figure}[tbp]
 \centerline{ {\psfig{figure=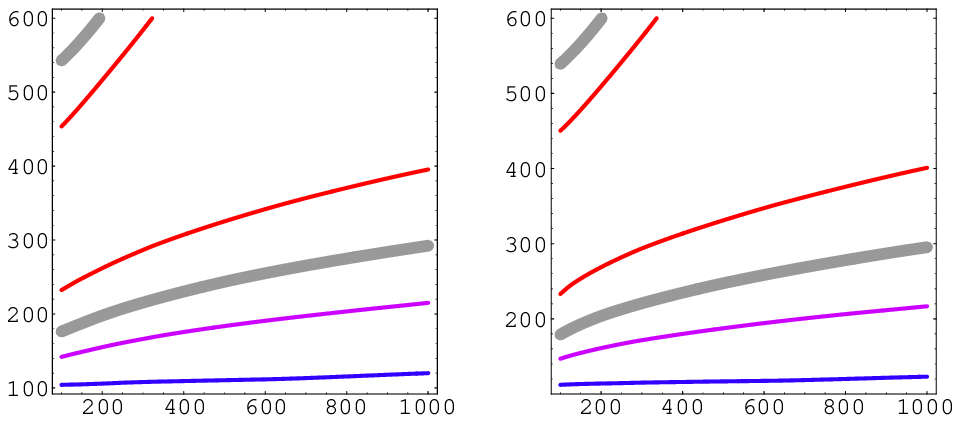,height=7.5cm}} }
  \begin{picture}(0,0)
\small
   \put(105,20){$m_{0}[\mbox{GeV}]$}
   \put(-14,124){$M_{1/2}[\mbox{GeV}]$}
   \put(330,20){$m_{0}[\mbox{GeV}]$}
   \put(80,220){${\rm{tan}}\beta=10,\;A_{0}=0$}
   \put(290,220){${\rm{tan}}\beta=30,\;A_{0}=300\GEV$}
   \put(125,132){${\bf{5\times10^{34}}}$}
    \put(65,170){${\bf{5\times10^{34}}}$}
   \put(167,82){${\bf{10^{35}}}$}
 \put(345,132){${\bf{5\times10^{34}}}$}
    \put(288,170){${\bf{5\times10^{34}}}$}
   \put(397,82){${\bf{10^{35}}}$}
\normalsize  
\end{picture}
 \caption{Contour plots of the proton life time in $m_{0}-M_{1/2}$
 plane for $\mu>0$ cases. Other conventions are the same as those in 
Fig.~\ref{fig:myunegative}.}
 \label{fig:myugyaku}
\end{figure}

We plot the contours of the life time of proton 
in $m_{0}-M_{1/2}$ plane, where $m_0$ and $M_{1/2}$ are the 
universal soft scalar mass and gaugino mass at the GUT scale, respectively.  
In Fig.\ref{fig:myunegative}, we show contours of the proton
life time for $\mu<0$ cases with several choices of 
$A_{0}\;(=0,\;-300\GEV)$, the universal A-term at the GUT scale, and
${\rm{tan}}\beta\;(=10,\;30)$.
The contour plots for $\mu>0$ cases are given in
Fig.~\ref{fig:myugyaku}.

As we can see from these contours, the proton life time is in the 
range $3\times 10^{34}-10^{35}$ yr. in most part of the parameter
space regardless of choices of ${\rm{tan}}\beta,\;A_{0}$ and 
sign of $\mu$.
We find the minimum of the proton life time is 
no less than $3\times 10^{34}$ yr. in whole parameter space,
which is well above the current experimental limit by the
Super-Kamiokande, $5.0 \times 10^{33}$ yr. (90\% C.L.)\cite{TITAND,SKdim6}. 
The thick gray contour lines  corresponding to the 
life time of proton $7\times 10^{34}$ yr. represent 
the $3\sigma$ discovery limit of the $1{\rm{Mt}}$ (fiducial volume)
detector after ten years running~\cite{HyperK,TITAND}.

Therefore, in the semi-simple unification model,
we have an intriguing possibility to confirm the
existence of the GUT in nature by observing the proton decay
in the next-generation ${\rm{Mt}}$ water $\check{\rm{C}}$erenkov detectors,
such as Hyper-Kamiokande~\cite{HyperK} and  TITAND~\cite{TITAND}.
In the optimistic cases with some enhancement factors of the decay rate of 
proton (see below), we have a chance to detect the proton decay
also in UNO~\cite{UNO} ($\sim 500{\rm{kt}}$ 
fiducial volume) experiment.

Although we set the cut-off scale $\Lambda$ to be $10^{17}\GEV$ in
calculating the GUT gauge boson mass to obtain the conservative 
lower bound of the proton decay rate, the actual cut-off scale 
may be a little more higher. 
In that case, the rate is enhanced by $(\Lambda/10^{17}\GEV)^2$.
Another possible enhancement of the decay rate arises when there 
are $\SU(5)_{\rm GUT}$-charged particles at an intermediate scale.
Existence of such particles are highly motivated in the semi-simple
unification model; ${\bf 5}+{\bf 5}^*$ representations are
required at the TeV scale when the discrete 
${\bf Z}_4$ R symmetry is gauged since the discrete gauge anomaly
${\bf Z}_{4 \rm R}$-[$\SU(5)_{\rm GUT}$]$^2$ should be
canceled~\cite{KMY}. In this case, 
the gauge coupling constant $\alpha_{\rm GUT}$ is stronger as a result
of the RG flow with new particles, and the decay rate is enhanced by
$1.6$. Although one might suspect that there is a one-loop threshold
correction from a possible mass splitting between triplets and doublets 
in ${\bf 5}+{\bf 5}^*$, and that the GUT gauge boson mass would be 
also changed, the GUT gauge boson mass is actually stable against this
correction, since Eq.(\ref{GUT-mass}) is an expression from which
the threshold corrections from the colored Higgs multiplets decouple.
The same thing happens when the SUSY breaking is
mediated through gauge mediation because of the presence of the 
messenger sector, though the SUSY threshold correction should be 
re-analyzed using the spectrum of the gauge mediated SUSY breaking 
in that case.

Finally, we summarize various uncertainties in the theoretical 
prediction given above.
The first uncertainty comes from possible violation of the 
$\U(3)_{\rm H}$ relation.  The violation $|(1/\alpha_{\rm
3H}-1/\alpha_{\rm 1H})(\Lambda)| = 1/3 $ leads to
a change in the decay rate by $\times/\div 0.5$. 
The second uncertainty comes from an error bar of the experimental 
values of the QCD coupling. 
This results in uncertainties by factor $ \times / \div 0.7$ for 
1$\sigma$ error.
The calculation of hadron matrix element in \cite{lattice} has 
an error  
$W = -0.153(19) \GEV^2$, which leads to a factor $\times / \div 0.8$.
Another uncertainty comes from a possible non-renormalization operators
involving the $\vev{\bar{Q}Q}$ vacuum expectation value
in the gauge kinetic function of the $\SU(5)_{\rm GUT}$\footnote{
Such non-renormalizable terms in the gauge kinetic
function is expected to be suppressed when one considers a certain
structure of the fundamental theory\cite{IWY}.} .
They generically modifies the $\SU(5)_{\rm GUT}$ relation 
directly by $\vev{\bar{Q}Q}/\Lambda^2 \simeq 10^{-2}$ at tree level. 
If it is the case, the possible change in the result will be at most
roughly the same amount as those discussed above.

There are two more sources of uncertainties whose effects we cannot 
estimate. 
First, if one considers an exotic situation in which unknown
non-renormalizable operators are relevant in the Wilsonian RG equations,
then the perturbative analysis we adopted in this letter is not 
adequate since we omitted such effects.
Secondly, we cannot estimate anything without the ${\cal N}=2$-SUSY relation.
This is because the perturbative analysis above the GUT scale 
is no longer valid without this relation, as is discussed in the appendix.

\section*{Acknowledgments}
Earlier part of this work was done in collaboration with K.Kurosawa.
The authors are grateful to Y.Shirman for discussion, to K.Suzuki 
for generating spectra of SUSY particles, and to T.Yanagida for 
discussion and careful reading of this manuscript.
M.F. and T.W. thank the Japan Society for the Promotion of Science for
financial support.

\appendix
\section{Role of approximate ${\cal N}=2$ SUSY relation \\
in perturbative analysis}

The GUT-breaking sector of the semi-simple unification model
has a multiplet structure of ${\cal N}=2$ SUSY, and 
the interactions between them (the first - the third lines 
in Eq.(\ref{eq:super}))
are quite similar to the ${\cal N}=2$ gauge interactions with 
Fayet-Iliopoulos F-term. Therefore, it is quite likely that 
this apparent ${\cal N}=2$ structure is a remnant of the 
${\cal N}=2$ SUSY in a fundamental theory\cite{IWY}. Then, 
the approximate ${\cal N}=2$ relation Eq.(\ref{eq:N=2-relation}) 
at the cut-off scale would be a natural consequence.

The approximate ${\cal N}=2$-relation is not only expected as above, 
but also almost required from another reason. The perturbation 
analysis performed in the text is no longer valid
if it is not satisfied and that is the reason why we assumed this
relation throughout this paper.

Let us suppose that the couplings $\lambda^{(')}_{\rm 3H}$ and 
$\lambda^{(')}_{\rm 1H}$ in the superpotential (\ref{eq:super}) 
are large compared with $g_{\rm 3H}$ and
$g_{\rm 1H}$. Then, 
those couplings become large extremely fast through one-loop RG equations, 
and hence we have to require that  
$\alpha^\lambda_{\rm 3H} \equiv \lambda_{\rm 3H}^2 /(4\pi)$ and 
$\alpha^\lambda_{\rm 1H} \equiv \lambda_{\rm 1H}^2 /(4\pi)$ are well below 
$2 \alpha_{\rm 3H}$ and $2 \alpha_{\rm 1H}$, respectively.
The same discussion also holds for $\lambda'_{\rm 3H}$ and
$\lambda'_{\rm 1H}$.
Now what if those couplings are small compared with the corresponding
gauge couplings?
In this case, we can neglect the last two terms in the following 
two-loop RG equations of the gauge couplings,
\begin{eqnarray}
\frac{\partial }{\partial \ln \mu}\left( \frac{1}{\alpha_{\rm 3H}}\right) 
& \simeq & \qquad  
- \frac{\alpha_{\rm 1H} + 17 \alpha_{\rm 3H}}{2 \pi^2}
+ \frac{\frac{5}{6}(\alpha^\lambda_{\rm 1H} + 17 \alpha^\lambda_{\rm 3H})}
       {2 \pi^2}
+ \frac{\frac{1}{6}(\alpha^{\lambda'}_{\rm 1H} + 17 \alpha^{\lambda'}_{\rm 3H})}
       {2 \pi^2},  \\
\frac{\partial }{\partial \ln \mu} \left( \frac{1}{\alpha_{\rm 1H}}\right) 
& \simeq  & -\frac{6}{2\pi} 
- \frac{\alpha_{\rm 1H} + 8 \alpha_{\rm 3H}}{2 \pi^2}
+ \frac{\frac{5}{6}(\alpha^\lambda_{\rm 1H} + 8 \alpha^\lambda_{\rm 3H})}
       {2 \pi^2}
+ \frac{\frac{1}{6}(\alpha^{\lambda'}_{\rm 1H} + 8 \alpha^{\lambda'}_{\rm 3H})}
       {2 \pi^2}.
\end{eqnarray}
Then, $\alpha_{\rm 3H}$ becomes large
quite rapidly and $\alpha_{\rm 1H}$ becomes large more faster than 
in the one-loop running. 
Thus, we require that $\alpha^{\lambda^{(')}}_{\rm 3H}$
and $\alpha_{\rm 1H}^{\lambda^{(')}}$ are 
comparable to the gauge couplings so that the two-loop effects are negligible.

In the approximate ${\cal N}=2$-SUSY limit and only in this limit,
$\alpha_{\rm 3H} \simeq \alpha_{\rm 3H}^{\lambda^{(')}}$ and 
$\alpha_{\rm 1H} \simeq \alpha_{\rm 1H}^{\lambda^{(')}}$, 
anomalous dimensions of hyper multiplets, 
\begin{equation}
\gamma_{Q_i} = 
 \frac{8 \alpha^{\lambda}_{\rm 3H} + \alpha^{\lambda}_{\rm 1H}}{6 \pi} 
- \frac{8 \alpha_{\rm 3H} + \alpha_{\rm 1H}}{6 \pi} + \cdots , 
\label{eq:anomalous-dim-hyp} \\
\end{equation}
vanish at all order, and the RG flows of the gauge couplings are one-loop exact.
Then, in turn, all other parameters in the superpotential, in particular 
$h$ and $h'$, are stable against quantum corrections from the strong
couplings $\alpha_{\rm 1H}$, $\alpha_{\rm 3H}$, 
$\alpha_{\rm 1H}^{\lambda^{(')}}$ and $\alpha_{\rm 3H}^{\lambda^{(')}}$.

Values of the coupling constants $h$ and $h'$ themselves are 
the possible obstruction left behind for the perturbative
analysis\footnote{We already know that other coupling constants
such as $y_{\bf 10,5^*}$ and $g_{\rm GUT}$ are weak and they are stable 
under their RG equations. Their perturbation to the approximate ${\cal
N}=2$-SUSY relation is also small enough.}.
They are obtained from a ratio $\sqrt{M_cM_{\bar{c}}}/M_v$, which in
turn is obtained from Eqs.(\ref{eq:full-matching3}-\ref{eq:full-matching1}) 
in the way described in the text:
\begin{equation}
\sqrt{hh'} = \sqrt{2}g_{\rm GUT} e^{\frac{\pi}{\alpha_{\rm 3H}}(\Lambda)}
\left(\frac{\Lambda}{\mu}\right)
\exp \left(\frac{2\pi}{12}
  \left( -\frac{4}{\alpha_3}+\frac{9}{\alpha_2}-\frac{5}{\alpha_1}\right)(\mu)
 \right)
\left(\frac{M_{8v}}{M_{8c}}\right)^2
e^{ \left(\frac{2\pi}{6}\left(\frac{1}{\alpha_{\rm 1H}}-\frac{1}{\alpha_{\rm 3H}}\right)(\Lambda)\right)} .
\end{equation}
The value of the RHS of this equation varies from sub-${\cal O}$(1) to
${\cal O}$(1).
Therefore, we can expect that the perturbative analysis performed 
in the text is valid for most part of the SUSY breaking parameter 
space, 
taking into account the uncertainties in the gauge coupling constants.


\newpage
\begin{table}
\begin{center}
\begin{tabular}{ccccccc}
\hline
Fields               & ${\bf 5}^*$,${\bf 10}$,${\bf 1}$ & $H$ , $\bar{H}$ &
                      $X^{\alpha}_{\;\beta} $ & $Q_i$,$\bar{Q}^i$ &
                                                $Q_6$  & $\bar{Q}^6$ \\
\hline
${\bf Z}_4$ R charge &                1               &     0           &
                             2                &      0            & 
                                                  2    &   -2        \\
\hline
\end{tabular}
\caption{Charge assignment of the ${\bf Z}_4$ R-symmetry is given. ${\bf 1}$ denotes a right handed neutrino.}
\label{tab:Z4-charge}
\end{center}
\end{table}
\begin{table}
\begin{center}
\begin{tabular}{ccccccc}
\hline
(${\bf 3},{\bf 2})^{-\frac{5}{6}}$ & 
$({\bf 3},{\bf 1})^{-\frac{1}{3}}$ &
$({\bf 3},{\bf 1})^{-\frac{1}{3}}$ &
$({\bf 1},{\bf 1})^0$ &
$({\bf 1},{\bf 1})^0$  &
$({\bf adj.},{\bf 1})^0$ &
$({\bf adj.},{\bf 1})^0$ \\
\hline
m.vect. & $\chi + \chi^{\dagger}$ & $\chi + \chi^{\dagger}$ &
m.vect. & $\chi + \chi^{\dagger}$ & m.vect. & $\chi + \chi^{\dagger}$ \\
\hline
$M_v = $ & $M_c =  $ & $M_{\bar{c}} =  $ &  $M_{1v} =  $ & $M_{1c} =  $ &
 $M_{8v} =  $ & $M_{8c} =  $ \\
$\sqrt{2}g_{\rm GUT}v$     &
$h v $   &  $h' v$    &
$\sqrt{2(g_{\rm 1H}^2 + 2g_{\rm GUT}^2/5)} v $  &
$\sqrt{2}\lambda_{\rm 1H} v $ &
$\sqrt{2(g_{\rm 3H}^2 + g_{\rm GUT}^2)} v $  &
$\sqrt{2}\lambda_{3H} v $  \\
\hline
\end{tabular}
\caption{Summary of the particle spectrum around the GUT scale. The
 first line denotes the representation under the MSSM gauge group. 
In the second line, m.vect. denotes ${\cal N}=1$ 
massive vector multiplet and $\chi + \chi^\dagger$ a pair of ${\cal N}=1$ 
chiral and anti-chiral multiplet. 
Mass of each multiplet is given in terms of gauge couplings
 and parameters in the superpotential (\ref{eq:super}) in the fourth line, 
and given in the third line is the expression of the mass used 
in the text. Multiplets with masses $M_{1v}$ and $M_{1c}$, $M_{8v}$ and
 $M_{8c}$ can be regarded as ${\cal N}=2$-SUSY partner with each other 
in the ${\cal N}=2$-SUSY limit(see also appendix).}
\label{tab:spec}
\end{center}
\end{table}
\end{document}